# THE PHYSICS AND IDENTITY OF DARK MATTER

TOM GEHRELS

Space Sciences Building, University of Arizona, Tucson, AZ 85721-0092

## ABSTRACT

This paper follows "The Physics and Identity of Dark Energy"; the identity is the acceleration energy of old photons. The present paper considers protons and everything else in the decay debris of our universe; it is an ensemble called "old protons, etc."

The accelerated expansion of our universe brings the debris into the inter-universal medium (IUM) of the multiverse, where it is conserved for long times, $\sim 10^{30}$ y, at $\sim 0$ K. Debris clouds eventually accrete from the IUM to grow into proto-universes. Special properties of decay and of old photons reduce the severe effects of gravity, and thereby avoid collapse into a black hole.

The beginning of our universe occurred when proton density of $10^{18}$ kg m$^{-3}$ was reached in a ~4.6 % central volume, at t $\sim 10^{-6}$ s on the standard-model clock; that is $10^{37}$ Planck times removed from a Big Bang. The Big Bang did not happen; the evolution of forces and subatomic components occurs in the IUM.

The protons, etc. are involved in the history of our universe as much as are the photons; they are the receivers of the photon acceleration. Dark matter therefore is the matter of "old protons, etc".

*Subject headings:* multiverse, universe, photons, protons, dark matter, expansion.

## 1. INTRODUCTION

The previous arXiv.1101.0161 paper, found dark energy to be the kinetic energy of acceleration of old photons, and the present paper explores the *recipients* of that acceleration. This study is possible only in a multiverse surrounding our universe (Gehrels 2007, 2011a, 2011b); the discoveries of the identities of dark energy and dark matter have not occurred until now because the multiverse was not considered.

Everything ages in our universe, and the decay debris floats on the accelerated expansion of intergalactic space into the inter-universal medium (IUM), in which our universe is embedded. New universes are accreted from IUM clouds that become proto-universes, and the old photons and old protons etc. are gravitationally re-energized. That is the beginning of a new universe like ours, $10^{37}$ Planck times removed from a Big Bang. Karl Schwarzschild (1916) had warned already that the Big Bang could not be, as is shown in the previous paper; he died in 1916 and his warning was not heeded.

The dark-energy scenario of the previous paper is described here again, but now is added what we know about dark matter etc. Section 2 spells out what is included in dark matter and in old protons. Section 3 has the beginning of our universe done again in two scenarios, for dark matter and for old protons. Again, a discovery occurs (Sec. 4). Section 5 summarizes the conclusions of the model and of both papers.

## 2. DARK MATTER

There is a large literature on the *puzzle* of dark matter; the public outreach for the Chandra X-ray spacecraft in 2009 summarizes that as follows.

"The name of "Dark Matter" is used to describe matter that can be inferred to exist from its gravitational effects, but does not emit or absorb detectable amounts of light. Observations of the rotational speed of spiral galaxies, the confinement of hot gas in galaxies and clusters of galaxies, the random motions of galaxies in clusters, the gravitational lensing of background objects, and the observed fluctuations in the cosmic microwave background radiation require the presence of additional gravity, which can be explained by the existence of dark matter.

The evidence suggests that the mass of dark matter in galaxies, clusters of galaxies, and the universe as a whole is about five or six times greater than the mass of ordinary light-emitting matter that makes up stars, planets, gas and dust. One possibility, considered unlikely by most astrophysicists, is that a modification of the theory of gravity can explain the effect attributed to dark matter.

The nature of dark matter is unknown. A substantial body of evidence indicates that it cannot be baryonic matter, i.e., protons and neutrons. The favored model is that dark matter is mostly composed of exotic particles formed when the universe was a fraction of a second old. Such particles, which would require an extension of the so-called Standard Model of elementary particle physics, could be WIMPs (weakly interacting massive particles), or axions, or sterile neutrinos.

Various types of experimental searches for dark matter candidates are being pursued by a number of investigators: the direct detection of dark matter particles using innovative new detectors; the detection of X-rays or gamma-rays from the decay or annihilation of dark matter particles; and the detection of dark matter particles created by colliding beams of high energy protons."

Padmanabhan (2002) concluded however "that *both* baryonic and nonbaryonic dark matter exist in the universe, with nonbaryonic being dominant." His general rule is that, "There is not a priori reason for the dark matter in different objects to be made of the same constituent." He discusses baryonic and nonbaryonic matter such as protons, WIMPs, axions, neutrinos, and massive astrophysical halo objects. For brevity, we shall use the term "protons etc.", and use it as

"Old cold protons and other particles such as old neutrons and old electrons are part of our universe's decay debris, as are old whole galaxies (each gravitationally holding its debris), old clusters of galaxies, and whatever other debris such as old stars; we refer to the ensemble as 'protons etc.' "

Interesting future work will be to sort out which of the members of the ensemble are basic, or are secondary ones of evolution within the universes.

The following section describes what happens when this protons etc. debris and that of the old cold photons accrete together to make our proto-universe. For interaction with the photons, the protons are of course of primary importance, and we recall that Andrei Sakharov pointed at the proton's long half-life, presently standing observationally at $10^{35}$ years, while he had derived that it might be $10^{50}$ or longer. With the present modeling, it becomes clear why there should be such large numbers.

## 3. THE HISTORY OF PROTONS ETC. FROM $10^{-6}$ s TO 380,000 y

Any multiverse would surround our universe; everything coming from our universe will be conserved in a multiverse; here we study the "Chandra Multiverse" described in detail in previous papers. The debris coming from decaying universes, via the inter-universal medium (IUM), and eventually into a new proto-universe, consists of old photons, but also of old protons etc. defined in the above four-lines quote.

Nothing stands still in the cosmos - clouds will grow, as they do in the interstellar medium (ISM), by sweeping material up during their motion through space; the material is the debris mixed from old universes all around. Eventually, self-gravitation becomes active, speeding the accretion of the cloud by its increasingly larger and effective gravitational cross-section.

The IUM *composition* is however totally different from that of the ISM; the decayed debris is energy-*seeking*, instead of the *active* atomic and molecular ISM material. The temperature in our growing proto-universe was therefore relatively constrained, to at most $10^{13}$ K (see below). The characteristic of energy-seeking is one of *two* participants in the solution to the classical problem that a cloud of mass equivalent to $10^{21}$ solar masses is much too large for accretion. However, the $10^{13}$-K temperature was apparently low enough for the mass concentrations of galactic clusters in our debris to survive, and to be recognized from the past in our present WMAP data (Hinshaw 2010). The other participant to that solution is *the acceleration pressure* of the photons, always moving at their high velocity c, and thereby providing essential outward counter-action to gravity so that the collapse to a black hole would not happen (this is discussed with Schwarzschild's warning in the previous paper).

Specifically for our universe, when the accretion in the *central* volume obtained the density of $10^{18}$ kg m$^{-3}$ - which is the density at which formation of photons and protons has been modeled in atomic theory - the re-energizing of the old photons became complete. Outside of the central volume, the re-energizing was incomplete because of insufficient density, but old photons remained active there in multiple scattering and expansion.

The central volume also had its old *protons* completely re-energized, a little tardier perhaps because there was more to it than for photons, and that might have given the photons a short time-spell for escape, to break out. A Photon Burst may then have been energetic enough to be observed by WMAP as a radiation signature with a *wider* curvature than that of the 3-K radiation (Hinshaw 2010). Characteristics for the epoch were $10^{18}$ kg m$^{-3}$, $10^{13}$ K, and t ~ $10^{-6}$ s, which is the epoch for formation of photons and protons in standard models.

If we re-define the beginning of our universe to have occurred at t = 0, it is t ~ $10^{-6}$ s on the Big-Bang clock, but neither the Big Bang nor any other of the early events happened (for t < $10^{-6}$ s) in our universe. They happened, and are happening today, but in the multiverse (not in the universes), which is the proper place for the evolution - with long times and many universes participating - of basic physics, forces, and subatomic particles. Their physics is that of h, c, G, H in the Chandra Multiverse. That is the truth, not just because Chandra's equations say so, but also for observational reasons: our h, c, G, H universe emerged from that multiverse, and the debris from our universe goes back into that multiverse.

After the above t ~ 10⁻⁶ s, the multiple scattering continued throughout the universe until age 380,000, when at space density of ~10⁻¹⁹ kg m⁻³ the electrons, protons, and neutrons combined to make atoms; they were hydrogen and helium atoms with wide internal spacing for the photons to escape through.

## 4. THE IDENTITY OF DARK MATTER

The protons interacted with the photons to cause multiple scattering, also by electrons and neutrons, slowing the photons' outward journey. Now we notice that the dark-energy actions for multiple scattering and expansion had a counterpart in the dark-matter actions of "protons etc." - the one gives, the other receives the kinetic energy of the photon acceleration.

In other words, while the dark energy is the kinetic energy of photons, dark matter appears to be the name for the ensemble with which the photons interact (defined as old protons etc. at the end of Sec. 2). What is therefore new in the present paper is that old protons etc. are the essential counterparts. The dark matter of old protons etc. is involved in the scenario as thoroughly as the dark energy of old photons. The reasoning of the previous paper is therefore extended to *dark matter being the old protons, etc.*

## 5. CONCLUSIONS

In summary of both papers, one sees the basic material of the cosmos that evolved at low temperatures in their ground states in the multiverse, and they seem be stored there to serve for more advanced evolutions in universes, stars, planets, and people.

The fundamental discovery of the present model is that the mass equation M(N) = (hc/G)$^{0.5N}$ shows that there are other universes, there is a multiverse. That is such a profound advance from our universe being the only one, and it is not surprising that the discoveries of dark-energy and dark matter were made. Other discoveries will be made with such a powerful model.

The dark-energy physics was a surprise because there had been so much speculation. The dark-matter physics should not be surprising because the acceleration of the photons *has* to interface with something. Actually, both solutions are simple and richly endowed with photons and protons etc. that age on an accelerated expansion and are conserved in the surrounding multiverse.

This dark-matter conclusion has the same simple but firm *confirmation* that we concluded for dark energy (*i.e.,* old photons). The evolution in the multiverse could not have left *unused* the major part of what all decaying universes contribute to the IUM, namely their 23% dark matter (*i.e.,* old protons, etc.). *If* these would have been left unused, the ever-increasing amount of dark matter (*i.e.* old protons, etc.) would have overwhelmed the cosmos. Our environment of matter and radiation is the supporting observation for the conclusions.